\documentclass{LMCS}
\usepackage[OT4]{fontenc}
\usepackage{amsmath}

\usepackage{macros}

\usepackage{xspace}
\usepackage{verbatim}
\usepackage{alltt}
\usepackage{latexsym}
\usepackage{enumerate}
\usepackage[pdftex]{hyperref}

\def\doi{4 (3:8) 2008}
\lmcsheading%
{\doi}
{1--20}
{}
{}
{Feb.~\phantom{0}1, 2007}
{Sep.~13, 2008}
{}

\begin{document}

\title[Consistency and Completeness of Rewriting in the Calculus of
  Constructions]{Consistency and Completeness of Rewriting in the Calculus of
  Constructions\rsuper*} 

\author[D.~Walukiewicz-Chrz{\k a}szcz]{Daria Walukiewicz-Chrz{\k a}szcz}
\address{Institute of Informatics,
    Warsaw University,
    ul. Banacha  2,
    02-097 Warsaw,
    Poland}
\email{\{daria,chrzaszcz\}@mimuw.edu.pl}
\thanks{This work was partly supported by Polish government grant
 3~T11C~002~27.}
\author[J.~Chrz{\k a}szcz]{Jacek Chrz{\k a}szcz}

\titlecomment{{\lsuper*}The extended abstract of this paper appeared earlier
as~\cite{chrzaszcze06ijcar}.}

\begin{abstract}
  Adding rewriting to a proof assistant based on the Curry-Howard
  isomorphism, such as Coq, may greatly improve usability of the
  tool. Unfortunately adding an arbitrary set of rewrite rules may
  render the underlying formal system undecidable and inconsistent.
  While ways to ensure termination and confluence, and hence
  decidability of type-checking, have already been studied to some
  extent, logical consistency has got little attention so far.

  In this paper we show that consistency is a consequence of
  canonicity, which in turn follows from the assumption that all
  functions defined by rewrite rules are complete.  We provide a sound
  and terminating, but necessarily incomplete algorithm to verify this
  property.
  The algorithm accepts all definitions that follow dependent
  pattern matching schemes presented by Coquand and studied by McBride in
  his PhD thesis. 
  It also accepts many definitions by rewriting including rules which
  depart from standard pattern matching.

%
\end{abstract}
%
%

\maketitle

\expandafter\ifx\csname SourceFile\endcsname\relax\else\SourceFile{intro.ltx}\fi\section{Introduction}

Equality is ubiquitous in mathematics. Yet it turns out that proof
assistants based on the Curry-Howard isomorphism, such as
Coq~\cite{coq}, are not very good at handling equality. While proving
an equality is not a problem in itself, using already established
equalities is quite problematic.
Apart from equalities resulting from internal reductions (namely, beta
and iota reductions),
which can be used via the conversion rule of the calculus of inductive
constructions without being recorded in
the proof term, any other use of an equality requires giving all
details about the context explicitly in the proof.
As a result, proof terms may become extremely large, taking up memory
and making type-checking time consuming: working with
equations in Coq is not very convenient.

A straightforward idea for reducing the size of proof terms is to
allow other equalities in the conversion, making their use
transparent. This can be done by using user-defined \emph{rewrite
  rules}. However, adding arbitrary rules may easily lead to logical
inconsistency, making the proof environment useless. It is of course
possible to put the responsibility on the user, but it is contrary to
the current Coq policy to guarantee consistency of developments
without axioms.  Therefore it is desirable to retain this guarantee when
rewriting is added to Coq. Since consistency is undecidable in
the presence of rewriting in general, one has to find some decidable criteria satisfied
only by rewriting systems which do not violate consistency.

The syntactical proof of consistency of the calculus of constructions,
which is the basis of the formalism implemented in Coq, requires every
term to have a normal form~\cite{DBarendregt:HandbookLiCS91}. The same
proof is also valid for the calculus of inductive
constructions~\cite{Dwerner94these}, which is even closer to the
formalism implemented in Coq.

There exist several techniques to prove (strong) normalization of the
calculus of constructions with
rewriting~\cite{Dbarbanera97jfp,Dblanqui99rta,blanqui03mscs,Ddaria2002,DWalukiewicz03},
following numerous works about rewriting in the simply-typed lambda
calculus.  Practical criteria for ensuring other fundamental
properties, like confluence, subject reduction and decidability of
type-checking are addressed e.g.  in~\cite{blanqui03mscs}.

Logical consistency is also studied in~\cite{blanqui03mscs}. It is
shown under the assumption that for every symbol \(f\) defined by
rewriting, \(f(t_1,\dots ,t_n)\) is reducible if \(t_1\dots t_n\) are terms
in normal form in the environment consisting of one type variable.
Apart from a proof sketch that this is the case for the two rules
defining the induction predicate for natural numbers and a remark that
this property resembles the completeness of definitions, practical ways to satisfy
the assumption of the consistency lemma are not discussed.

Techniques for checking completeness of definitions are
known for almost 30 years for the first-order algebraic setting
\cite{Dguttag78,Dthiel84,Dkounalis85}. More recently, their adaptations
to type theory appeared in
\cite{DCoquand92typesa,mcbride00dependently} and
\cite{schurmann03coverage}. In this paper we show how the latter
algorithm can be tailored to the calculus of constructions extended
with rewriting. We study a system where the set of available function
symbols and rewrite rules are not known from the beginning but may grow as
the proof development advances, as it is the case with concrete
implementations of modern proof assistants.


We show that logical consistency is an easy consequence of canonicity,
which in turn can be proved from completeness of definitions by
rewriting, provided that termination and confluence are proved first.
Our completeness checking algorithm closes the list of necessary
procedures needed to guarantee logical consistency of developments in
a proof assistant based on the calculus of constructions with
rewriting. 

In fact, in this paper we work in a framework which is slightly more
general than the calculus of constructions, namely that of pure type
systems, of which the calculus of constructions is an instance.
However, since termination and confluence are used both in our
algorithm and in the proof of its correctness, our results are useful
only if a termination and confluence criteria exist for a given pure
type system extended with rewriting. Some work in this direction has
been done, e.g., in~\cite{DBarthe:1997:TAT}.

\section{Rewriting in the Calculus of Constructions}%
\label{sec:rewriting}
Let us briefly
discuss how we imagine introducing rewriting in Coq and
what problems we encounter on the way to a usable system.

From the user's perspective definitions by rewriting could be entered
just as all other definitions:\footnote{The syntax of the definition
  by rewriting is inspired by the experimental ``recriture'' branch of
  Coq developed by Blanqui. For the sake of clarity we omit certain
  details, like environments of rule variables and allow the infix +
  in the definition.}
\begin{program}
Inductive  nat : Set  :=  O : nat | S : nat \ra nat.
Symbol + : nat \ra nat \ra nat
Rules
  O + y \lra  y
  x + O \lra  x
  (S x) + y \lra S (x + y)
  x + (S y) \lra S (x + y)
  x + (y + z) \lra (x + y) + z.
Parameter n : nat.
\end{program}
The above fragment 
can be interpreted as an environment consisting of the inductive
definition of natural numbers, symmetric definition by rewriting of
addition and the declaration of a variable $n$ of type \nat. In this
environment all rules for + contribute to conversion. 
For~instance both $\forall x\!:\!\nat. \; x+0 =x $ and
$\forall x\!:\!\nat.\; 0+x=x $ can be proved by $\lam{x}{\nat}\,
\mathit{refl}\;\nat \;x$, where $\mathit{refl}$ is the only constructor of
the Leibniz equality inductive predicate.
Note that the definition of + is terminating and confluent. The latter
can be checked by an (automatic) examination of its critical pairs.

Rewrite rules can also be used to define higher-order and polymorphic
functions, like the \texttt{map} function on polymorphic lists.
In this example, the first two rules correspond to the usual
definition of \texttt{map} by pattern matching and structural
recursion and the third rule can be used to quickly get rid of the
\texttt{map} function in case one knows that \texttt{f} is the
identity function. 
\begin{program}
Symbol map : forall (A:Set), (A \ra A) \ra list A \ra list A
Rules
  map A f (nil A) \lra nil A
  map A f (cons A a l) \lra cons A (f a) (map A f l)
  map A (fun x \(\Rightarrow\) x) l \lra l
\end{program}

Even though we consider higher-order rewriting,
we choose the simple matching modulo $\alpha$-conversion.
Higher-order matching
is useful for example to encode logical languages by higher-order
abstract syntax, but it is seldom used in Coq where modeling relies rather
on inductive types. 
Instead of higher-order matching, one needs 
a possibility not to specify certain arguments in left-hand sides, and hence 
to work with rewrite rules built from terms that may be not typable.
Consider, for example the type {\tt tree} of trees with size, holding
some Boolean values in the nodes, and
the function {\tt rotr} performing a right rotation in the root of the
tree.

\label{sec:rotr}
\begin{program}
Inductive tree : nat \ra Set :=
  Leaf : tree O
| Node : forall n1:nat, tree n1 \ra bool \ra forall n2:nat, tree n2 
            \ra tree (S(n1+n2)).

Symbol rotr : forall n:nat, (tree n) \ra (tree n)
Rules
   rotr 0 t  \lra  t
   rotr ? (Node O t1 a n2 t2)  \lra  Node O t1 a n2 t2
   rotr ?1 (Node ?2 (Node ?3 A b ?4 C) d ?5 E)  
             \lra  Node ?3 A b (S (?4 + ?5))(Node ?4 C d ?5 E) 
\end{program}
The first argument of {\tt rotr} is the size of the tree and the
second is the tree itself. The first two rules cover the trees which
cannot be rotated and the third one performs the rotation. 

The ?\ marks above should be treated as different variables.
The information they hide is redundant for typable terms: if we take
the third rule for example, the values of ?3, ?4 and ?5 must correspond
to the sizes of the trees A, C and E respectively, ?2 must be equal to
S(?3+?4) and ?1 to S(?2+?5). Note that by not writing these subterms
we make the rule left-linear (and therefore easier to match) and avoid
critical pairs with +, hereby helping the confluence proof.

This way of writing left-hand sides of rules was already used by Werner
in~\cite{Dwerner94these} to define elimination rules for inductive
types, making them orthogonal (the left-hand sides are of the form 
$I_{elim}\;P\;\vec{f}\;\vec{w}\;(c\; \vec{x})$, where $P$, $\vec{f}$,
$\vec{w}$, $\vec{x}$ are distinct variables and $c$ is a
constructor of $I$).
In~\cite{blanqui03mscs}, Blanqui gives a precise
account of these omissions using them to make more rewriting rules
left-linear. 
Later, the authors of~\cite{brady04types} show that these
redundant subterms can be completely removed from terms (in a calculus
without rewriting however). In~\cite{Gregoire-Barras-05}, a new optimized
convertibility test algorithm is presented for Coq, which ignores
testing equality of these redundant arguments.

In our paper we do not explicitly specify which arguments should/could
be replaced by~?\ and do not restrict left-hand sides to be
left-linear.  Instead, we rely on an acceptance condition to suitably
restrict the form of acceptable definitions by rewriting to guarantee
the needed metatheoretical properties listed in the next section.

It is also interesting to note that when the first argument of {\tt
  rotr} is ?1 then we may understand it as S(?2+?5) matched to terms
modulo the convertibility relation and not just syntactically (i.e.,
modulo $\alpha$-conversion).

\section{Pure Type Systems with Generative Definitions}%
\label{sec:pts}

Even though most papers motivated by the development of Coq
concentrate on the calculus of constructions, we present here a
slightly more general formalization of a pure type system with
inductive definitions and definitions by
rewriting. The presentation, taken
from~\cite{chrzaszcz03types,chrzaszcz04phd}, is quite close to the way
these elements could be implemented in Coq. The
formalism is built upon a set of PTS sorts \PTSSorts, a binary
relation \PTSAxioms and a ternary relation \PTSRules over \PTSSorts
governing the typing rules \mreg{Term/Ax} and \mreg{Term/Prod}
respectively (Figure~\ref{fig:rules}).  The syntactic class of pseudoterms
is defined as follows:
\begin{center}
  $t$ ::= $v$ $|$ $\sOrt$ $|$ $t_1\ t_2$ 
                $|$ $\lam{v}{t_1}t_2$ $|$ $\prd{v}{t_1}t_2$
\end{center}
A pseudoterm can be a variable $v \in \Var$, a sort \(\sOrt \in \PTSSorts\), an
application, an abstraction or a product.
We write \(|t|\) to denote the size of the pseudoterm $t$, with
\(|v|=|\sOrt|=1\).
We use Greek letters $\gamma,\delta$ to denote substitutions
which are finite partial maps from variables to pseudoterms. The
postfix notation is used for the application of substitutions to
pseudoterms. 

Inductive definitions and definitions by rewriting are
\emph{generative}, i.e.\ they are stored in the
environment and are used in terms only through names
they ``generate''. An environment is a
sequence of declarations, each of them is
a variable declaration $v:t$, an inductive definition \mbox{$\ind$}, where
$\Gamma^I$ and $\Gamma^C$ are environments providing names and types of (possibly mutually
defined) inductive types and their constructors, or a definition by
rewriting $\rew$, where $\Gamma$ is an environment providing names and types of (possibly
mutually defined) function symbols and $R$ is a set of rewrite rules
defining them. Types of inductive types, constructors and function
symbols determine their arity: given $v:t$ in an inductive
definition or a definition by rewriting, if $t$ is of the form 
\(\prd{x_1}{t_1}\dots \prd{x_n}{t_n}\hat{t}\) where \(\hat{t}\) is not a
product, then $n$ is the arity of $v$.

A rewrite rule is a triple denoted by \(\Delta  \vdash l \longrightarrow r\), where $l$
and $r$ are pseudoterms and \(\Delta \) is an environment, providing
names and types of variables occurring in the left- and right-hand
sides $l$ and $r$.

Given an environment $E$, inductive types, constructors
and function symbols declared in $E$ are called constants (even though
syntactically they are variables). We often write \(h(e_1,\dots ,e_n)\)
to denote the application of a constant $h$ to pseudoterms $e_1$,
\ldots, $e_n$, when $n$ is the arity of $h$.
General environments are denoted by
$E$ and environments containing only variable declarations are
denoted by $\Gamma$, $\Delta$, $G$, $D$.  We assume that names of all
declarations in environments are pairwise disjoint. A pair consisting
of an environment $E$ and a term $e$ is called a sequent and denoted
by \(E \vdash e\). A sequent is well-typed if \(E \vdash e : t\) for some $t$.

\begin{defi}\label{def:pts}
  A \emph{pure type system with generative definitions} is defined by
  the
  typing rules in Figure~\ref{fig:rules}, where:
  \begin{enumerate}[$\bullet$]
  \item \Pos\ is a positivity condition for inductive definitions (see
    assumptions below).
  \item \Acc\ is an acceptance condition for definitions by rewriting
    (\textit{idem}).
  \item The relation $\approx$ used in the rule \mreg{Term/Conv} is
    the smallest congruence on well typed terms, generated by \(%
 \longrightarrow\) which is the sum of beta and rewrite
    reductions, denoted by \( \longrightarrow_{\beta}  \) and \( \longrightarrow_{R}  \) respectively
    (for exact definition see~\cite{chrzaszcz04phd}, Section
    2.8).

\item The notation $\delta:\Gamma \ra E$ means that $\delta$ is a
  \emph{well-typed substitution}, i.e.\ $E \vdash
  v\delta:t\delta$ for all  $v:t \in \Gamma $.

\end{enumerate}

\end{defi}
\begin{figure}[t]
  \framebox{
  \begin{minipage}{1\textwidth}
    \centering
    Let\ \ $\Gamma^I=\jenv{I}{t^I}{n}$ \ \ and \ \ 
    $\Gamma^C=\jenv{c}{t^C}{m}$\\[2ex]
  \begin{math}
      \displaystyle
      \frac      {
        \begin{array}{c}
          E          \vdash t^I_j:\sOrt_j   \qquad   
                     t^I_j=\overrightarrow{(z:Z_j)}\,\sOrt'_j   \qquad
                     \textrm{ for } j=1\dots n \\[0.5ex]
          E;\Gamma^I \vdash t^C_i:\hat{\sOrt}_i  \qquad 
                     t^C_i=\overrightarrow{(z:Z'_i)}\,I_{j_i}\vec{w} \qquad
                     \textrm{ for } i=1\dots m \\
        \end{array}
      }
      {   E \vdash \Ind(\Gamma^I:=\Gamma^C) : \correct
      }
      \textrm{\ \ if\ } \Pos_E(\Gamma^I:=\Gamma^C)
  \end{math}
\\[1ex]
\linia\\[1ex]
Let\ \ 
$\Gamma = \jenv{f}{t}{n}$ \ \ and \ \ 
$R=\{\Gamma_i : l_i \lra r_i\}_{i=1\dots m}$, where
$\Gamma_i=x^i_1:t^i_1;\dots;x^i_{n_i}:t^i_{n_i}$
\\[1ex]
      \begin{math}      
      \displaystyle\frac{
   \begin{array}{c}
      E \vdash t_k : \sOrt_k \quad \textrm{for } k=1\dots n\\[0.1ex]
      
      E;\Gamma_i \vdash \ok \quad \mathit{FV}(l_i,r_i) \subseteq \Gamma_i \quad
      \textrm{for } i=1\dots m
   \end{array}
   }
   {   \begin{array}{c}E\vdash \Rew(\Gamma, R) : \correct\end{array}
   }
   {\textrm{\ \ if\ }\Acc_E(\Gamma,R)}
    \end{math}
  \end{minipage}}

  \framebox{
    \begin{minipage}{1\textwidth}
      \centering
      \begin{math}
        \begin{array}{ccc}
          \displaystyle\frac{}{\eps \vdash \ok}
          &&
          \displaystyle
          \frac{ 
            E \vdash \ok 
            \quad
            E \vdash t:\sOrt
          }{
            E; v:t \vdash \ok
          }
          \\[3ex]
          \displaystyle
          \frac{ 
            E \vdash \ok 
            \quad
            E \vdash \ind : \correct
          }{
            E; \ind \vdash \ok
          }
          &\qquad&
          \displaystyle
          \frac{ 
            E \vdash \ok 
            \quad
            E \vdash \rew : \correct
          }{
            E; \rew \vdash \ok
          }
        \end{array}
      \end{math}  
    \end{minipage}}

  \framebox{
  \begin{minipage}{1\textwidth}
  \centering
  \begin{math}
    \begin{array}{c}
      \displaystyle
      \frac{
        E_1;v:t;E_2 \vdash \ok
      }{
        E_1;v:t;E_2 \vdash v:t
      }
    \end{array}
  \end{math}
  \\[0.5ex]

  \begin{math}
    \begin{array}{ccccc}
      \displaystyle
      
      \frac{
        E \vdash \ok
      }{
        E \vdash I_i : t^I_i
      }

      &\quad&

      \displaystyle
      \frac{
        E \vdash \ok
      }{
        E \vdash c_i:t^C_i
      }        

      &\quad\text{where}&
      
      \left\{
        \begin{array}{l}
          E = E_1; \ind; E_2
          \\
          \Gamma^I = \jenv{I}{t^I}{n}
          \\
          \Gamma^C = \jenv{c}{t^C}{m}
        \end{array}
      \right.

      \\[4ex]
      \displaystyle
      \frac{
        E \vdash \ok
      }{
        E \vdash f_i : t_i
      }
      &\quad&
      \displaystyle
      \frac{
        E \vdash \ok
        \quad
        \delta : \Gamma_i \ra E
      }{
        E \vdash l_i\delta \lra_R r_i\delta
      }
      &\quad\text{where }&
      \left\{
      \begin{array}{l}
        E=E_1;\rew;E_2
        \\
        \Gamma = \jenv{f}{t}{n}
        \\
        R = \{\Gamma_i : l_i \lra r_i \}_{i=1\dots m}
     \end{array}
      \right.
    \end{array}
  \end{math}
  \end{minipage}}

  \framebox{
  \begin{minipage}{1\textwidth}
  \centering
   \begin{math}
    \wregula{Term/Prod}{
      E \vdash t_1 : \sOrt_1
      \quad
      E; v:t_1 \vdash t_2 : \sOrt_2
    }
    {
      E \vdash \prd{v}{t_1} t_2 : \sOrt_3
    } 
    {\text{\ where\ } \sOrt_1, \sOrt_2, \sOrt_3 \in \PTSSorts}
    \ \ \
    \nregula{Term/Abs}{
      E; v:t_1 \vdash e:t_2 
      \quad
      E \vdash \prd{v}{t_1} t_2 : \sOrt
    }
    {
      E \vdash \lam{v}{t_1} e : \prd{v}{t_1} t_2
    } 
    \ \ \
    \wregula{Term/Ax}  
    {
      E \vdash \ok
    }
    {
      E \vdash \sOrt_1 : \sOrt_2
    } 
    {\text{where}\ (\sOrt_1,\sOrt_2) \in \PTSAxioms}
  \end{math}
  \\[1ex]
  \begin{math}
    \nregula{Term/App}
    {
      E \vdash e : \prd{v}{t_1} t_2
      \quad
      E \vdash e' : t_1
    }
    {
      E \vdash e\; e' : t_2\{v \mapsto e'\}
    } 
    \quad
    \nregula{Term/Conv}
    {
      E \vdash e : t
      \quad
      E \vdash t' : \sOrt
      \quad
      E \vdash t \approx t'
    }
    {
      E \vdash e : t'
    }
  \end{math}
\end{minipage}}
  \caption{Definition correctness, environment correctness and lookup, PTS rules}
  \label{fig:rules}
\end{figure}


\noindent 
As in~\cite{DWalukiewicz03,blanqui03mscs}, recursors and their
reduction rules have no special status and they
are supposed to be expressed by rewriting. 

\paragraph{\bf Assumptions.} 
We assume that we are given a positivity condition \Pos\ for inductive
definitions and an acceptance condition
\Acc\ for definitions by rewriting.  Together with the right choice of
the PTS they must imply the following properties:
\begin{enumerate}[\bf P1]
\item subject reduction, i.e.\ \(E \vdash e:t\), \(E \vdash e \longrightarrow e'\) implies
  \(E \vdash e' : t\)
\item uniqueness of types, i.e.\ \(E \vdash e:t\), \(E \vdash e:t'\) implies 
  \(E \vdash t \approx  t'\).
\item strong normalization, i.e.\ \(E \vdash \ok \) implies that
  reductions of all well-typed terms in \(E\) are finite
\item confluence, i.e.\ \(E \vdash e:t\), \(E \vdash e \longrightarrow^* e'\),
  \(E \vdash e \longrightarrow^* e''\) implies \(E \vdash e' \longrightarrow^* \hat{e}\) and \(E \vdash e''
  \longrightarrow^* \hat{e}\) for some $\hat{e}$.

\end{enumerate}
These properties are usually true in all well-behaved type theories.
They are for example all proved for the calculus of algebraic
constructions~\cite{blanqui03mscs}, an extension of the calculus of
constructions with inductive types and rewriting, where \Pos\ is the strict positivity
condition as defined in~\cite{DPM:inddsc-c}, and \Acc\ is the General
Schema.

From now on, we use the notation \(t \norm \) for the unique normal form of $t$.

\section{Consistency and Completeness}

Consistency of the calculus of constructions (resp. calculus of
inductive constructions) can be shown by rejecting all cases of 
a hypothetical normalized proof $e$ of \(\prd{x}{*}x\) in a \emph{closed}
environment, i.e.\ empty environment (resp. an environment   
containing only inductive definitions and no axioms). Our goal is to
extend the definition of closed environments to the calculus of
constructions with rewriting, allowing it to include 
a certain class of definitions by rewriting.

Let us try to identify that class. If we reanalyze $e$ in the new
setting, the only new possible normal form of $e$ is an application
$f(\vec{e})$ of a function symbol $f$, coming from a rewrite
definition $\Rew(\Gamma, R)$, to some arguments in normal form. There
is no obvious argument why such terms cannot be proofs of
\(\prd{x}{*}x\). On the other hand if we knew that such terms were
always reducible, we could complete the consistency proof. Let us call
\comp{} the condition on rewrite definitions we are looking~for (i.e.\
$f(\vec{e})$ is always reducible), which can also be read as: the
function symbols from $\Gamma$ are completely defined by the set
of rules $R$.

Note that the completeness of $f$ has to be
checked much earlier than it is used: we use it in a given closed environment
\(E=E_1;\Rew (\Gamma, R); E_2\) but it has to be checked when
$f$ is added to the environment, i.e.\ in the environment $E_1$. It
implies that completeness checking has to account for environment
extension and can be performed only with respect to arguments of such
types, of which the set of normal forms could not change in the future.
This is the case for arguments of inductive types.

The requirement that functions defined by rewriting are completely
defined could very well be included in the condition \Acc. On the
other hand, the separation between \Acc\ and \Comp\ is motivated by
the idea of working with abstract function symbols, equipped with some
rewrite rules not defining them completely. For example if
+ from Section~\ref{sec:rewriting} were declared using only
the third rewrite rule, one could develop a theory of an
associative function over natural numbers.



The intuition behind the definitions given below is the following.
A rewrite definition \(\Rew(\Gamma ,R)\) is \emph{complete}
(satisfies \(\Comp(\Gamma,R)\)) if for all $f$
in $\Gamma$, the goal \(f(x_1,\dots ,x_n)\) is \emph{covered} by $R$. A
goal is covered if all its instances are \emph{immediately
 covered}, i.e.\ head-reducible by $R$. Following
the discussion above we limit ourselves to
\emph{normalized canonical instances}, i.e.\ built from
constructors wherever possible.

\begin{defi}[Canonical form and canonical substitution]
\label{def:CANONICAL}
  Given a judgment \(E \vdash e:t\) we say that the term $e$ is in
  \emph{canonical} form if and only if:
  \begin{enumerate}[$\bullet$]
  \item if \(t \norm \) is an inductive type then \(e=c(e_1,\dots ,e_n)\) for
    some constructor $c$ and terms \(e_1,\dots ,e_n\) in canonical form
  \item otherwise $e$ is arbitrary
  \end{enumerate}
Let $\Delta$ be a variable environment and $E$ a correct
environment. We call $\delta: \Delta \ra E$  \emph{canonical} if for
every variable $x\in\Delta$, the term $x\delta$ is canonical.
\end{defi}
From now on, let $E$ be a global environment and let
\(\Rew (\Gamma,R)\) be a rewrite definition such that \(%
E \vdash \Rew (\Gamma,R) : \correct \). Let \mbox{\(f:(x_1:t_1)\dots (x_n:t_n)\,t \in
\Gamma\)} be a function symbol of arity $n$.
\begin{defi}\label{def:GOAL}
  A \emph{goal} is a well-typed sequent 
  \(E;\Gamma;\Delta \vdash f(e_1,\dots ,e_n)\).

  A \emph{normalized canonical instance} of the goal 
  \(E;\Gamma;\Delta \vdash f(e_1,\dots ,e_n)\)
  is a well-typed sequent 
  \(E;\Rew (\Gamma ,R);E' \vdash f(e_1\delta \norm ,\dots ,e_n\delta \norm )\) for any
  canonical substitution  \(\delta: \Delta \rightarrow  E;\Rew (\Gamma ,R);E'\).

  A term $e$ is \emph{immediately
    covered} by $R$ if there is a rule \(G \vdash l \longrightarrow r\)
  in $R$ and a substitution~\(\gamma\) 
  such that
  $l\gamma = e$. 
  By obvious extension we can also write that a goal or a normalized
  canonical instance is immediately covered~by~$R$.

  A goal is \emph{covered} by $R$ if all its normalized canonical
  instances are immediately covered by $R$.
\end{defi}
Note that, formally, a normalized canonical instance is not a goal. The
difference is that the conversion corresponding to the environment of
an instance contains reductions defined by $R$, while the one of
a goal does~not.
\begin{defi}[Complete definition]
\label{def:CERTIFIED DEFINITION} 
  A rewrite definition \(\Rew (\Gamma ;R)\) is \emph{complete} in the
  environment $E$, which is denoted by \(\Comp_E(\Gamma;R)\), if and
  only if for all function symbols \(f:(x_1:t_1)\dots (x_n:t_n)\,t \in
  \Gamma\) the goal \(E;\Gamma;x_1:t_1;\dots ;x_n:t_n \vdash f(x_1,\dots ,x_n)\)
  is covered by~$R$.
\end{defi}

\begin{exa}
  The terms 
  \texttt{(S O)}, 
  \texttt{$\lambda$x:nat.x} and
  \texttt{(Node O Leaf true O Leaf)} are canonical, while 
  \texttt{(O + O)} and 
  \texttt{(Node nA A b O Leaf)} are not.
  Given the definition of \texttt{rotr} from Section~\ref{sec:rotr} 
  consider the following terms:
\begin{program}
  \(t\sb{1}=\;\)\texttt{rotr (S (nA + nC)) (Node nA A b nC C)}
  \(t\sb{2}=\;\)\texttt{rotr (S O) (Node O Leaf true O Leaf)}
\end{program}
Both (with their respective environments) are goals for
\texttt{rotr}, and $t_2$ (with a slightly different
environment) is also a
normalized canonical instance of $t_1$.  The goal $t_1$ is
not immediately covered, but its instance $t_2$ is, as it is
head-reducible by the second rule defining \texttt{rotr}.
Since
other instances of $t_1$ are also immediately covered, the goal is
covered (see Example~\ref{ex:alg}).
\end{exa}

\label{sec:consistency}

It follows that completeness of definitions by rewriting
guarantees canonicity and logical consistency.  

\begin{defi}\label{def:CLOSED ENVIRONMENT} 
  An environment $E$ is \emph{closed} if and only if it contains only
  inductive definitions and complete definitions by rewriting, i.e.\
  for each partition of $E$ into \(E_1;\Rew (\Gamma, R);E_2\) the
  condition $\Comp_{E_1}(\Gamma, R)$ is satisfied.
\end{defi}

\begin{lem}[Canonicity]\label{lem:CLOSED DAJE CANONICAL}
Let $E$ be a closed environment. If \(E \vdash e:t\) and $e$ is in normal
form then $e$ is canonical.
\end{lem}
\begin{proof}
By induction on the size of $e$. 
If \(t \norm  \) is not an inductive type then any $e$ is canonical.
Otherwise, let us analyze the structure of $e$. It cannot be a
product, an abstraction or a sort because \(t \norm \) is an inductive
type. Since $E$ is closed, it is not a variable either. 
Hence $e$ is of the form \(e' e_1 \dots  e_m\) (with $m$ possibly equal
0), where $e'$ is not an application.
The term
$e'$ can be neither a product, nor a sort (they cannot be applied), nor a variable
($E$ is closed). It is not an abstraction, since $e$ is in normal
form. The only possibility left is that $e'$ is a constant $h$ of
arity $n\leq m$, and we
get \(e=h(e_1,\dots ,e_n)\;e_{n+1} \dots  e_m\).

  
Since \(t \norm  \) is an inductive type, $h$ cannot be an inductive
type. If it is a constructor then $n=m$ and 
by induction hypothesis \(e_1,\dots ,e_n\)
are in canonical form and so is \(h(e_1,\dots ,e_n)\).
If $h$ is a function symbol then
\(E=E_1;\Rew (\Gamma ,R);E_2\) for some \(E_1, E_2\) and \mbox{$h:(x_1:t_1)
\ldots (x_n:t_n)\,\hat{t} \in \Gamma$} of arity $n\leq m$.  
Since $E$ is closed, \(\Rew (\Gamma ,R)\) is complete. Let us
show that \(E \vdash  h(e_1,\dots ,e_n)\) is a normalized canonical instance of
\(E_1; \Gamma ; \Delta  \vdash  h(x_1, \ldots ,x_n)\), where \(\Delta  =
x_1:t_1;\dots ;x_n :t_n\).  By 
induction hypothesis, terms $e_1, \ldots e_n$ are canonical and
consequently \(\delta : \Delta  \rightarrow  E\) defined by \mbox{$\delta(x_i)
= e_i$} is canonical. Moreover, \mbox{\(h(e_1,\dots ,e_n) = h(x_1\delta \norm  ,
\dots ,x_n\delta \norm  )\)} since $e_1, \ldots e_n$ are in normal form.
But every normalized canonical instance of a complete definition is
reducible, which contradicts the assumption that
\(e=h(e_1,\dots ,e_n)\;e_{n+1} \dots  e_m\)
is in normal form.
\end{proof}

\begin{thm}\label{thm:closed is consistent}
  Every closed environment is consistent.
\end{thm}

\begin{proof}
  Let $E$ be a closed environment. Suppose that 
  \(E \vdash e : \prd{x}{\star}x\).  Since \(E \vdash \ok \) and 
  \(E \vdash \Ind (\False :\star:=) : \correct \) we have \(E' \vdash \ok \) where
  \(E'=E;\Ind (\False :\star:=)\). Moreover $E'$ is a closed environment.
  
  Hence, we have \(E' \vdash e\; \False  : \False \). By Lemma~\ref{lem:CLOSED
    DAJE CANONICAL}, the normal form of \(e\;\False \) is canonical. 
  Since \(\False \) has no constructors, this is impossible.
\end{proof}

\section{Checking Completeness}

The objective of this section is to provide an algorithm for
checking completeness of definitions by rewriting. The algorithm
presented in Subsection~\ref{sec:algorithm} checks that a goal is
covered using successive \emph{splitting}
(Definition~\ref{def:SPLITTING ALONG x}), i.e., replacement of
variables of inductive types by constructor patterns. In order to know
which constructor terms can replace a given variable, one has to
compare types and hence an algorithm for unification modulo conversion
is needed (Definition~\ref{def:CORRECT UNIFICATION ALGORITHM}).
Consider for example the first rule of the definition of
\texttt{rotr}. It is clear that only \texttt{Leaf} can replace
\texttt{t} in \texttt{rotr O t} because other trees have types that do
not unify with \texttt{tree O}.

Correctness of the completeness checking algorithm is proved in
Lemma~\ref{lem:CORRECTNESS OF THE ALGORITHM}. It is done using an
additional assumption on rewrite systems called \emph{preservation of
  reducibility} which is discussed in Subsection~\ref{sec:preservation}.


\begin{defi}[Unification problem]
A quadruple \(E,\Delta \vdash t \unifeq  s\), where $E$ is an environment, 
$\Delta$ a variable environment and $s$, $t$ are terms,   is a
\emph{unification  equation in E}. A~\emph{unification problem in E}
is a finite set of unification equations. Without loss of generality
we may assume that the variable environments $\Delta$ in all equations
are the same.  

A \emph{unifier} or a \emph{solution} of the unification problem $U$
is a substitution \(\gamma: \Delta \rightarrow  E;E'\) such
that \(E;E' \vdash t\gamma \approx s\gamma\) 
for every \(E,\Delta \vdash t \unifeq  s\) in~$U$. We say that \(E'\) is the
co-domain of~$\gamma$, which is denoted by \(Ran(\gamma)\).

A unifier \(\gamma\) is the \emph{most general
  unifier} if \(Ran(\gamma)\) is a variable environment $\Delta'$ and
for every unifier \(\delta: \Delta \rightarrow  E;E''\) there
exist a substitution \(\delta':\Delta ' \rightarrow  E;E''\),
such that \(E;E'' \vdash \delta \approx  \gamma;\delta'\). 
\end{defi}

\begin{defi}[Correct unification algorithm]
\label{def:CORRECT UNIFICATION ALGORITHM}
  A \emph{unification algorithm} is a procedure which 
  for every unification problem \(U=\{E,\Delta \vdash t_i \unifeq  s_i\}\) returns a
  substitution \(\gamma \), a bottom $\bot$, or a question mark $?$.
  The algorithm is \emph{correct} if and only if: if it answers~\(\gamma \), it is
  the most general unifier \(\gamma: \Delta \rightarrow  E;\Delta'\) 
  such that $\Delta' \subseteq \Delta$ and for all \(x \in \Delta'\), 
  \(\gamma(x)=x\); if it answers~$\bot$, $U$~has no unifier. 
\end{defi}

Since unification modulo conversion is undecidable, every correct
unification algorithm must return ?\ in some cases, which may be seen
as too difficult for the algorithm.  An example of
such a partial unification algorithm is constructor unification, that
is first-order unification with constructors and type constructors as
rigid symbols, answering $?$ whenever one compares a non-trivial pair
of terms involving non-rigid symbols.

From now on we assume the existence of a correct (partial) unification
algorithm \(Alg\).


\begin{defi}[Splitting]\label{def:SPLITTING ALONG x}
  Let \(E;\Gamma;\Delta \vdash f(\vec{e})\) be a goal. A
  variable $x$ is a \emph{splitting variable} if   $x:t \in
  \Delta$ and \(t \norm  = I\vec{u}\) for  some inductive type $I \in E$. 

  A \emph{splitting operation} considers all constructors $c$ of the
  inductive type $I$ and for each of them constructs the following
  unification problem $U_c$:
\begin{displaymath}%
E;\Gamma, \; \Delta; \Delta_c \vdash x \unifeq  c(z_1, \ldots z_k)
\qquad
\quad
E; \Gamma, \; \Delta; \Delta_c \vdash I\vec{u} \unifeq  I\vec{w}\end{displaymath}
where $c:(z_1:Z_1)\ldots (z_k:Z_k).I\vec{w}$ and $\Delta_c =
z_1:Z_1,\ldots, z_k:Z_k$. 

If for all constructors $c$, \(Alg(U_c) \not =\;  ?\), 
the splitting is \emph{successful}. 
In that case, let
\(Sp(x)=\{\sigma_c\;|\; \sigma_c = Alg(U_c) \wedge Alg(U_c) \not = \bot\}\). 
The result of splitting is the set of goals
\(\{E;\Gamma;Ran(\sigma_c) \vdash f(\vec{e})\sigma_c\}_{\sigma_c \in Sp(x)}\).

If \(Alg(U_c) =\; ?\) for some $c$, the splitting \emph{fails}.

\end{defi}
\begin{exa}\label{ex:splitting}
  If one splits the goal \texttt{rotr n t} along 
  \texttt{n}, one gets two goals: \texttt{rotr O t} and \texttt{rotr
    (S m) t}. The first one is immediately covered by the first
  rule for \texttt{rotr} and if we split the second one along
  \texttt{t}, the \texttt{Leaf} case is impossible, because
  \texttt{tree O} does not unify with \texttt{tree (S~m)} 
  and the \texttt{Node} case gives \texttt{rotr (S (nA + nC)) (Node nA A b nC
    C)}.
\end{exa}
The following lemma states the correctness of splitting, i.e.\ that
splitting does not decrease the set of normalized canonical instances.
Note that the lemma would also hold if we
had a unification algorithm returning an arbitrary set of most
general solutions, but in order for the coverage checking algorithm to
terminate the set of goals resulting from splitting must be finite.

\begin{lem}
\label{NORMALIZED CANONICAL INSTANCES AFTER SPLITTING}
  Let \(E;\Gamma;\Delta \vdash f(\vec{e})\) be a coverage goal and let
  \(\{E;\Gamma;Ran(\sigma_c) \vdash f(\vec{e})\sigma_c\}_{\sigma_c \in Sp(x)}\)
  be the result of
  successful splitting along $x:I\vec{u} \in \Delta$. Then every
  normalized canonical instance of \(%
E;\Gamma;\Delta \vdash   f(\vec{e})\) is a normalized canonical instance of
  \(E;\Gamma; Ran(\sigma_c) \vdash f(\vec{e})\sigma_c\) for some $\sigma_c
  \in Sp(x)$.  
\end{lem}
\begin{proof}
  Let \(E;\Rew (\Gamma ,R);E' \vdash f(e_1\delta \norm ,\dots ,e_n\delta \norm )\)
  \mar{Dlaczego $Rew(\Gamma)$ a nie $\Gamma$}be a normalized canonical
  instance according to a substitution \(\delta : \Delta \rightarrow 
  E;\Rew (\Gamma ,R);E'\). Since $\delta$ is canonical, $x\delta$ is a
  constructor term $c(s_1, \ldots s_k)$ for some constructor
  $c:(z_1:Z_1)\ldots (z_k:Z_k).I\vec{w}$ of $I$. Let us show that
  \(E;\Rew (\Gamma ,R);E' \vdash f(e_1\delta \norm ,\dots ,e_n\delta \norm )\) is a
  normalized canonical instance of \(%
E;\Gamma; Ran(\sigma_c) \vdash   f(e_1\sigma_c,\dots ,e_n\sigma_c)\). Let $\Delta_c = z_1:Z_1,\ldots,
  z_k:Z_k$.

  First note that \(\delta \cup [\vec{s}/\vec{z}] : \Delta;\Delta_c \rightarrow 
  E;\Rew (\Gamma ,R);E'\) is a solution of the unification problem
  \(E;\Gamma,\, \Delta; \Delta_c \vdash x \unifeq  c(z_1, \ldots z_k)\) and \(E;
  \Gamma, \, \Delta; \Delta_c \vdash I\vec{u} \unifeq  I\vec{w}\), from the
  definition of splitting. Indeed, $x\delta=c(s_1, \ldots s_k)=c(z_1,
  \ldots z_k)[\vec{s}/\vec{z}]$ and \(E;\Rew (\Gamma ,R);E' \vdash (I\vec{u})
  \delta \approx (I \vec{w}) [\vec{s}/\vec{z}]\) since they are both
  types of $c(s_1, \ldots s_k)$ in \(E;\Rew (\Gamma ,R);E'\).

  By definition of $\sigma_c$, which is the most general unifier
  computed by a correct unification algorithm, \(E;\Gamma;Ran(\sigma_c)
  \vdash \delta \cup [\vec{s}/\vec{z}] \approx  \sigma_c;\delta'\) for some
  \(\delta':Ran(\sigma_c) \rightarrow  E;\Rew (\Gamma ,R);E'\), where
  $Ran(\sigma_c) \subseteq \Delta;\Delta_c$. Consequently,
  \(E;\Gamma;Ran(\sigma_c) \vdash \delta'(z_m) \approx  s_m\) for $z_m \in
  \Delta_c$ and \(E;\Gamma;Ran(\sigma_c) \vdash \delta'(y) \approx  \delta(y)\)
  for $y \in \Delta$. Since $\vec{s}$ are canonical terms \(\delta'\norm \)
  is a canonical substitution.

Let us look closely at \(%
E;\Rew (\Gamma ,R);E' \vdash f((e_1\sigma_c)(\delta'\norm )\norm ,\dots ,(e_n\sigma_c)(\delta'\norm )\norm )\) which is a
normalized \(\delta' \norm \)-instance of \(%
E;\Gamma; Ran(\sigma_c) \vdash f(e_1\sigma_c,\dots ,e_n\sigma_c)\). Since \(E;\Gamma;Ran(\sigma_c) \vdash \linebreak[4]
\delta \cup [\vec{s}/\vec{z}] \approx  \sigma_c;\delta'\), we have
\((e_m\sigma_c)(\delta'\norm )\norm  = (e_m\sigma_c\delta')\norm  = (e_m\delta)\norm \) for
every $m$. Consequently, \(E;\Rew (\Gamma ,R);E' \vdash f(e_1\delta
\norm ,\dots ,e_n\delta \norm )\) is a normalized canonical instance of \(E;\Gamma;
Ran(\sigma_c) \vdash f(e_1\sigma_c,\dots ,e_n\sigma_c)\).
\end{proof}

\subsection{Preservation of Reducibility}\label{sec:preservation}

Although one would expect that an immediately covered goal is also
covered, it is not always true, even for confluent systems. It turns out that 
we need a property of critical pairs that is stronger than just
joinability.  Let us suppose that \texttt{or\;:\;bool \ra\ bool \ra\ bool} is
defined by four rules by cases over \texttt{true} and \texttt{false}
and that \texttt{if\;:\;bool \ra\ bool \ra\ bool \ra\ bool} is defined by two
rules by cases on the first argument.
 
\begin{program}
Inductive I : bool \ra Set := C : forall b:bool, I (or b b).
Symbol f:  forall b:bool, I b \ra bool
Rules
  f (or b b) (C b) \lra if b (f true (C true)) (f false (C false))
\end{program}
In the example presented above all expressions used in types and rules
are in normal form, all critical pairs are joinable, the system is
terminating, and splitting of \texttt{f b i} along \texttt{i} results
in the only reducible goal \texttt{f (or b b) (C b)}. In spite of that
\texttt{f} is not completely defined, as \texttt{f true (C true)} is a
normalized canonical instance of \texttt{f (or b b) (C b)} and it is
not reducible.
In order to know that an immediately covered goal is always covered we
need one more condition on rewrite rules, called preservation of
reducibility.
\begin{defi}
\label{PRESERVATION OF REDUCIBILITY}
Definition by rewriting \(\Rew (\Gamma, R)\) \emph{preserves
  reducibility} in an environment $E$ if for every critical pair 
$\langle f(\vec{u}), r\delta \rangle$ of a rule \(G_1 \vdash f(\vec{e}) \longrightarrow r\)
in $R$ with a rule \(G_2 \vdash g \longrightarrow d\) coming from $R$ or from some other
rewrite definition in $E$, the term \(f(\vec{u} \norm  )\) is
head-reducible by~$R$.
\end{defi}
Note that by using ?\ variables in rewrite rules one can get rid of
(some) critical pairs and hence make a definition by rewriting satisfy
this property. In the example above one could write \texttt{f ?\ (C b)}
as the left-hand side. This would also make the system non-terminating,
and show that $f$ is not really well-defined.  

Of course all orthogonal rewrite systems,
in particular inductive elimination schemes, as defined
in~\cite{Dwerner94these}, preserve reducibility.

\begin{lem}\label{lem:REDUCIBILITY} 
Let \(E \vdash e:t\) and \(e=f(e_1, \ldots e_n)\), where $f$ of arity $n$
comes from
\(\Rew (\Gamma,R)\) which preserves reducibility. If $e$ is
head-reducible by $R$ then \(f(e_1 \norm  , \ldots e_n \norm  )\) is also
head-reducible by~$R$.
\end{lem} 
\begin{proof}
By induction on \( \longrightarrow \). 
If $e_1, \ldots e_n$ are in normal forms then the conclusion is
obvious. Otherwise, let  \(G_1 \vdash f(\vec{l}) \longrightarrow r\) be a rule from $R$
and $\gamma$  a substitution such that $f(\vec{e})=
f(\vec{l})\gamma$ and let us make one reduction step \(e_i \longrightarrow e_i'\),
using the rule \(G_2 \vdash g \longrightarrow d\).

There are two possibilities: the reduction in $e_i$ happens either in
substitution $\gamma$, i.e.\ in the term $\gamma(x)$, where $x$ is a
free variable of $f(\vec{l})$, or it happens on a position $p$ that
belongs to~$f(\vec{l})$. In the former case, let us do identical
reduction in all other instances of $x$. Obviously, we get a term
\(f(e'_1, \ldots e'_n)\) that is smaller than $e$ in \( \longrightarrow \) and
is still an instance of $f(\vec{l})$. Hence by induction hypothesis we
get the desired conclusion.

Otherwise, $f(\vec{l})$ and $g$ superpose at some nonvariable position 
and we have $f(\vec{l})|_p\gamma =g \xi$ for some position $p$ and substitution
$\xi$.  Since we may suppose that free variables of $f(\vec{l})$ and
$g$ are different, we get $f(\vec{l})|_p(\gamma \cup \xi) =g (\gamma
\cup \xi)$. Let $\delta$ be the most general unifier of $f(\vec{l})|_p$ and
$g$ and let $\langle f(\vec{u}), r\delta \rangle$ be the corresponding
critical pair. Since $\delta$ is the most general unifier, there
exists $\sigma$ such that $(\gamma \cup \xi) = \delta;\sigma$ and
$f(\vec{e}) = f(\vec{l})\gamma = f(\vec{l})(\gamma \cup \xi) =
f(\vec{l})\delta\sigma$ with \(f(\vec{l})\delta\sigma \rightarrow _R
f(\vec{u})\sigma = f(e_1, \ldots e_i' \ldots e_n)\). 
By preservation of reducibility \(f(\vec{u} \norm  )\) is head-reducible by
$R$. Hence \(f(\vec{u} \norm  )\sigma\) is also head-reducible by $R$. Like
above we can apply induction hypothesis and deduce that \(f(\vec{e}
\norm )\) is head-reducible by $R$.
\end{proof}

\begin{lem}
\label{IMMEDIATELY COVERED IS COVERED} 
Let \(\Rew (\Gamma, R)\) preserve reducibility in an environment $E$,
let $f \in \Gamma$ and
let  \(E;\Gamma ;\Delta \vdash f(\vec{e})\) be a goal. If it is immediately covered
then it is covered. 
\end{lem}
\begin{proof}
Let \(E;\Gamma ;\Delta \vdash f(\vec{e})\) be a goal immediately covered
by $R$ and \(\delta :\Delta \rightarrow  E; \Rew (\Gamma,R); E'\) be a
canonical substitution. Obviously, \(E; \Rew (\Gamma,R); E' \vdash f(\vec{e}
\delta)\) is immediately covered by~$R$. Hence, by
Lemma~\ref{lem:REDUCIBILITY} 
\(E; \Rew (\Gamma,R); E' \vdash f(\vec{e} \delta \norm  )\) is also immediately
covered by $R$, i.e.\ \(E;\Gamma ;\Delta \vdash f(\vec{e})\) is covered.
\end{proof}

\subsection{Coverage Checking Algorithm}\label{sec:algorithm}

In this section we present an algorithm checking whether a set of
goals is covered by the given set of rewrite rules. The algorithm is
correct only for definitions that preserve reducibility. The
algorithm, in a loop, picks a goal, checks whether it is immediately
covered, and if not, splits the goal replacing it by the subgoals
resulting from splitting.  In order to ensure termination, splitting is
limited to \emph{safe splitting variables}.  Intuitively, a splitting
variable is safe if it lies within the contour of the left-hand side
of some rule when we superpose the tree representation of the
left-hand side with the tree representation of the goal.
The number of nodes that have to be added to the goal in
order to fill the tree of the left-hand side is called a distance,
and a sum of distances over all rules is called
a measure.
Since the measures of goals resulting from splitting
are smaller than the measure of the original goal,
the coverage checking algorithm terminates.

This subsection is organized as follows. We start by defining the
splitting matching algorithm which is used to define safe splitting
variables. Next, we provide definitions and lemmas needed to prove
termination of the coverage checking algorithm and then we give the
algorithm itself and the proof of its correctness. We conclude this
subsection with some positive and negative examples leading to an
extension of the algorithm allowing us to accept definitions by case
analysis even if the unification algorithm is not strong enough.

Let us start with the splitting matching algorithm which finds
variables in $t_1$ that lie within the contour of $t_2$.

\begin{defi}[Splitting matching]
  The \emph{splitting matching algorithm} is defined in
  Figure~\ref{fig:splitmatch}.  Given two sequents \(\Delta_1\vdash t_1\)
  and \(\Delta_2\vdash t_2\), it returns the unique set $S$, such that
  \(t_1 <_\Lambda t_2 \Rightarrow  S\) is derivable.  The set $S$ is a subset of
  $\{\bot\} \cup \{p \in Pos(t_1)\; |\; t_1|_p \in \Delta_1\}$.
\end{defi}
\begin{figure}[t]
  \framebox{
    \begin{minipage}{1\textwidth}
      \centering
$$
\begin{array}{ccc}\displaystyle{}%
\frac{t_1 <_{p\cdot 1} t_1' \Rightarrow  S_1\qquad  t_2 <_{p\cdot 2} t_2' \Rightarrow   S_2}{\prd{x}{t_1}{t_2} <_p \prd{x}{t_1'}{t_2'} \Rightarrow  S_1 \cup S_2}&\qquad&
\displaystyle{}%
\frac{t_1 <_{p\cdot 1} t_1' \Rightarrow  S_1\qquad  t_2 <_{p\cdot 2} t_2' \Rightarrow   S_2}{\lam{x}{t_1}{t_2} <_p \lam{x}{t_1'}{t_2'} \Rightarrow  S_1 \cup S_2}\end{array}
$$
$$%
\frac{t_1 <_{p\cdot 1} t_1' \Rightarrow  S_1\qquad t_2 <_{p\cdot 2} t_2' \Rightarrow  S_2}{\app{t_1}{t_2} <_p \app{t_1'}{t_2'} \Rightarrow  S_1 \cup S_2}$$
$$%
\frac{t_1 <_{p\cdot 1} t_1' \Rightarrow  S_1\quad \dots \quad t_n <_{p\cdot n} t_n' \Rightarrow  S_n}{h(t_1,\dots ,t_n) <_p h(t_1',\dots ,t_n') \Rightarrow  S_1 \cup \dots  \cup S_n}\ h \textrm{ is a constant}     $$
$$
\begin{array}{ccc}\displaystyle{}%
\frac{t_1, t_2 \not \in \Var \quad  \head(t_1)\not= \head(t_2)}{t_1 <_p t_2 \Rightarrow  \{\bot\}}&\qquad&
\displaystyle{}%
\frac{t_1 \in \Delta_1\quad  t_2 \not \in (\Var  \setminus \Delta_2)}{t_1 <_p t_2 \Rightarrow  \{p\}}\\[3ex]
\displaystyle{}%
\frac{\begin{array}{c}(t_1 \in (\Var  \setminus \Delta_1) \wedge t_1=t_2) \vee\\
(t_1 \not \in \Var  \wedge t_2 \in \Delta_2)\end{array}}{t_1 <_p t_2 \Rightarrow  \emptyset}&\qquad&
\displaystyle{}%
\frac{\begin{array}{c}(t_1 \in (\Var  \setminus \Delta_1) \wedge t_1\not=t_2) \vee\\
(t_2 \in (\Var  \setminus \Delta_2) \wedge t_1\not=t_2)\end{array}}{t_1 <_p t_2 \Rightarrow  \{\bot\}}\end{array}
$$
\end{minipage}}
\caption{Splitting matching rules, parametrized by $\Delta_1$, $\Delta_2$}\label{fig:splitmatch}
\end{figure}


\begin{defi}[Safe splitting variable]
\label{def:SAFE SPLITTING VARIABLE}
Let \(\Delta_1 \vdash t_1\) and \(\Delta_2 \vdash t_2\) be sequents such that
$t_2$ is a left-hand side of a rule from $R$ and let $S$ be a set such
that \(t_1 <_\Lambda t_2 \Rightarrow  S\) and $\bot \not \in S$. A variable $x
\in \Delta_1$ is a \emph{safe splitting variable for \(%
\Delta_1 \vdash   t_1\) along \(\Delta_2 \vdash t_2\)} if it is a splitting variable and
there exists $p \in S$ such that $t_1|_p=x$ and either $t_2|_p$ is a
variable declared in $\Delta_2$ or $t_2|_p= c(\vec{e})$ for some
constructor $c$ and some terms $\vec{e}$.

The set of safe splitting variables for the sequent \(\Delta_1 \vdash t_1\) along
\(\Delta_2 \vdash t_2\) is denoted by \mbox{\(%
SV(\Delta_1 \vdash t_1, \Delta_2 \vdash t_2)\)} or \(SV(t_1,t_2)\) for short. \(SV(t,R)\) is the
set of safe splitting variables for $t$ along left-hand sides of rules
from $R$.  
\end{defi}
\begin{exa}\label{ex:safe-variables}
  In the goal \texttt{rotr (S (S nC)) (Node O Leaf b (S nC) C)} there
  are two safe splitting variables \texttt{b} and \texttt{C} along
  the left-hand sides of the rules defining \texttt{rotr}.
\end{exa}
\begin{defi}[Distance]
\label{def:DISTANCE}
Let \(\Delta_1 \vdash t_1\) and \(\Delta_2 \vdash t_2\) be sequents and $S$ be
a set such that \(t_1 <_\Lambda t_2 \Rightarrow  S\). If $\bot \not \in  S$ then
the  \emph{distance} of \(\Delta_1 \vdash t_1\) from \(\Delta_2 \vdash t_2\),
denoted by \(dist(\Delta_1 \vdash t_1, \Delta_2 \vdash t_2)\) or
\(dist(t_1,t_2)\), equals 
$\sum_{p \in S} |t_2|_p|$; otherwise it is equal to $0$. 
\end{defi} 
\noindent
The following two lemmas state that the distance of a term decreases
when we apply a substitution, and it decreases strictly if it is a
substitution resulting from splitting. 

\begin{lem}[Distance of a substituted sequent]
\label{lem:DISTANCE AFTER SUBSTITUTION}
Let \(\Delta_1 \vdash t_1\) and \(\Delta_2 \vdash t_2\) be sequents and let $S$ be
a set such that \(t_1 <_\Lambda t_2 \Rightarrow  S\). Then for every substitution
\(\gamma: \Delta_1 \ra \Delta'\) we have 
\(dist(\Delta' \vdash t_1\gamma, \Delta_2 \vdash t_2) \leq dist(\Delta_1 \vdash t_1,
\Delta_2 \vdash t_2)\). 

Moreover, if $\bot \in S$ then \(dist(\Delta' \vdash t_1\gamma, \Delta_2
\vdash t_2) = dist(\Delta_1 \vdash t_1, \Delta_2 \vdash t_2) = 0\). 
\end{lem}
\begin{proof}
Let $S_\gamma$  be a set such that \(t_1\gamma <_\Lambda t_2 \Rightarrow  S_\gamma\) and let
us denote  \(dist(\Delta_1 \vdash t_1, \Delta_2 \vdash t_2)\) by~$d$ and
\(dist(\Delta' \vdash t_1\gamma, \Delta_2 \vdash t_2)\) by $d_\gamma$.

If $\bot \in S$ then \(d=0\). Note that $\bot \in S$ if and only if
there is a position $p$ such that  subterms occurring at $p$ in $t_1$
and $t_2$ either  have different head symbols, or $t_2|_p$
(resp. $t_1|_p$) is a bound variable in $t_2$ (resp. $t_1$) and
$t_1|_p\not = t_2|_p$.
Of course, if we compare $t_1\gamma|_p$ and $t_2|_p$ then 
either they still have different head-symbols or $t_2|_p$
(resp. $t_1|_p$) is a bound variable and $t_1\gamma|_p\not =
t_2|_p$. Hence \(d_\gamma = 0\). 

If $\bot \not \in S$ then \(d= \sum_{p \in S} |t_2|_p| \geq 0\). If $\bot
\in S_\gamma$ then obviously $0= d_\gamma \leq d$. Otherwise, let us take $p
\in S$ and the set $Q_p = \{q \in S_\gamma \; | \; p \prf q \}$, where
$\prf$ is the prefix ordering.  Since all positions from $Q_p$ are
independent (as \(t_1\gamma|_q \in \Var \) for every $q \in S_\gamma$) we have
$\sum_{q \in Q_p} |t_2|_q| \leq |t_2|_p|$ and the equality holds only
if $Q_p=\{p\}$.  Let us show that $\forall q\! \in\! S_\gamma \;\exists p\! \in\! S
\; p \prf q$.  Indeed, assuming that $\bot \not \in S_\gamma$, $q \in S_\gamma$
either because $q \in S$ and $(t_1|_q)\gamma \in \Delta'$ or because
there is a position $p \in S$ such that $q=p\cdot q'$ for some $q'$
and $(t_1|_p\gamma)|_{q'} \in \Delta'$. Of course, 
since positions in $S$ are independent, the sets $Q_p$ are disjoint for different $p$.

Hence $S_\gamma= \bigcup_{p \in S} Q_p$ and   
$d_\gamma = \sum_{q \in S_\gamma} |t_2|_q| = \sum_{p \in S} \sum_{q \in Q_p}
|t_2|_q| \leq \sum_{p \in S} |t_2|_p|=d$.
\end{proof}

\begin{lem}[Distance after splitting strictly decreases]
\label{lem:DISTANCE AFTER SPLITING DECREASES}
Let \(E;\Gamma;\Delta \vdash f(e_1,\dots ,e_n)\) be a goal,
\(t=f(e_1,\dots ,e_n)\), let
\(G \vdash l \longrightarrow r\) be one of the rewrite rules for $f$ in $R$  
and let $S$ be a set such that \(t <_\Lambda l  \Rightarrow  S\)
and $\bot \not \in S$. If  $x:I\vec{u} \in SV(t,l)$ is a safe splitting
variable and splitting $t$ along $x$ is successful then 
\(dist(Ran(\sigma_c) \vdash t\sigma_c,G \vdash l) < dist(\Delta \vdash t,G \vdash l)\)
for every $\sigma_c \in Sp(x)$.  
\end{lem}
\begin{proof}
Let $\sigma_c\in Sp(x)$ and let $S_c$ be a set such that
\(t\sigma_c<_\Lambda l\Rightarrow  S_c\). By Lemma~\ref{lem:DISTANCE AFTER
SUBSTITUTION} we have 
\(dist(t\sigma_c,l) \leq  dist(t,l)\). Let us analyze the proof of
that lemma and show that in case of a substitution resulting from
splitting there is a strict inequality between \(dist(t\sigma_c,l)\)
and \(dist(t,l)\). In the proof it was noticed that for every
$p \in S$, $\sum_{q \in Q_p} |l|_q| \leq |l|_p|$, where $Q_p=\{q \in
S_c \;|\; p \prf q\}$ and that $\sum_{q \in
  Q_p} |l|_q| = |l|_p|$ only 
if $Q_p=\{p\}$. Consequently, if we show that there exists a position $p$
such that  $p \not \in Q_p$,  we immediately get
\(dist(t\sigma_c,l) < dist(t,l)\).      

Since $x:I\vec{u}\,$\ is a safe splitting variable for $t$ along $l$,
there exists a position $p \in S$ such that $t|_p =x$ and $l|_p \in G$
or $l|_p=c'(\vec{a})$ for some constructor $c'$. Since $\sigma_c$
results from successful splitting, $x\sigma_c = c(\vec{b})$ for some
$\vec{b}$. Now, there are three cases. If $l|_p \in G$ then
\(t\sigma_c<_\Lambda l\Rightarrow  \emptyset\), $Q_p= \emptyset$ (and hence $p
\not \in Q_p$). Otherwise, if $c'\not = c$ then we fall into an easy
case when $\bot \in S_c$ and \(dist(t\sigma_c,l)=0 < |c'(\vec{a})|
\leq dist(t,l)\). Finally, if $c'=c$, the computation of $S_c$ passes
through the step \(c(\vec{b}) <_p c(\vec{a}) \Rightarrow  \_\). This means that
all positions in $Q_p$ come from~$\vec{b}$ and that they are longer
than $p$ or that $Q_p=\emptyset$. Thus $p \not \in Q_p$ and \(dist(t\sigma_c,l)< dist(t,l)\).
\end{proof}

\begin{defi}[Measure of a goal]
\label{def:MEASURE}
 Let \(E;\Gamma;\Delta \vdash f(e_1,\dots ,e_n)\) be a goal and let \(R_f= \{
 G_i \vdash l_i \longrightarrow r_i\}_{i=1\dots m}\) be the set of rules for
 $f$. The \emph{measure of \(E;\Gamma;\Delta \vdash f(e_1,\dots ,e_n)\)}
 equals \(\sum_{i=1\dots m} dist(\Delta \vdash f(e_1,\dots ,e_n), G_i \vdash l_i)\).
\end{defi} 
It follows directly from Lemmas~\ref{lem:DISTANCE AFTER SUBSTITUTION}
and~\ref{lem:DISTANCE AFTER SPLITING DECREASES} that the measure of a
goal strictly decreases after applying a substitution resulting from
splitting.

\begin{lem}[Measure after splitting strictly decreases]
\label{lem:MEASURE AFTER SPLITING DECREASES}
 Let \(E;\Gamma;\Delta \vdash f(e_1,\dots ,e_n)\) be a goal,
 $t=f(e_1,...,e_n)$,   \(R_f= \{ G_i \vdash l_i \longrightarrow r_i\}_{i=1\dots m}\) 
 be the set of rules for $f$ and
 $S$ be a set such that \(t <_\Lambda l_j \Rightarrow  S\) for some \(j \in \{1,\dots ,m\}\)
 and $\bot \not \in S$.  If $x:I\vec{u} \in SV(t,R)$ is a safe splitting
 variable and \(\{\phi_1,\dots ,\phi_n\}\) is the result of successful
 splitting of $t$ along x then the measure of every \(\phi_i\) is
 strictly smaller than the measure of $t$.
\end{lem}
\begin{proof}
For every $\sigma_c \in Sp(x)$, we have to show that $\sum_{i=1..m} dist(t\sigma_c,l_i)<
\sum_{i=1..m} dist(t,l_i)$. This follows from
\(dist(t\sigma_c,l_j) <  dist(t,l_j)\),
which is the consequence of Lemma~\ref{lem:DISTANCE AFTER SPLITING DECREASES}
and \(dist(t\sigma_c,l_i) \leq dist(t,l_i)\) for all \(i = 1\dots m\),
$i\not = j$, which follows from  Lemma~\ref{lem:DISTANCE AFTER
  SUBSTITUTION}. 
\end{proof}

\begin{defi}[Coverage checking algorithm]
\label{def:COVER CHECKING ALGORITHM}
Let $W$ be a set of pairs consisting of a goal and a set of safe
variables of that goal along left-hand sides of rules from $R$ and
let $CE$ be a set of goals.
The coverage checking algorithm works as follows:\\[1ex]
Initialize\\
\(\mbox{}\ \ W=\{(E;\Gamma;x_1:t_1;\dots ;x_n:t_n \vdash f(x_1,\dots ,x_n),\ \  
                         SV(f(x_1,\dots ,x_n),R))\}\)\\
\(\mbox{}\ \ CE=\emptyset\)\\[1ex]
Repeat \vspace{-1ex}
\begin{enumerate}[(1)]
\item choose a pair $(\phi, X)$ from $W$,
\item if $\phi$ is immediately covered by one of the rules from $R$ then\\
   $W:=W \setminus \{(\phi,X)\}$ 
\item otherwise
  \begin{enumerate}[(a)] 
  \item if $X=\emptyset$ then $W:=W \setminus \{(\phi,X)\}$, $CE:=CE
     \cup \{\phi\}$ 
  \item otherwise choose $x \in X$; split $\phi$ along $x$
     \begin{enumerate}[(i)]
     \item if splitting is successful and returns
       \(\{\phi_1,\dots ,\phi_n\}\) then \\
       $W:=W\setminus \{(\phi,X)\} \cup \{(\phi_i, SV(\phi_i, R))\}_{i=1...n}$,
     \item otherwise $W:=W \setminus \{(\phi,X)\} \cup \{(\phi, X \setminus \{x\})\}$
     \end{enumerate}
  \end{enumerate}
\end{enumerate}
until $W=\emptyset$
\end{defi}

\begin{lem}
\label{lem:TERMINATION}
The cover checking algorithm terminates.
\end{lem}
\begin{proof}
  Let us consider the following measure $M(W)$: the multiset of
  lexicographically ordered pairs consisting of the measure of $\phi$ and
  the size of $X$, for all $(\phi,X) \in W$. We will show that every loop
  of the algorithm strictly decreases $M(W)$. Consider $(\phi, X) \in W$.
  If $\phi$ is immediately covered then obviously the measure of
  $W\setminus \{(\phi,X)\}$ is strictly smaller than the measure of $W$.
  Otherwise, we split $\phi$ along some $x \in X$. If splitting fails then
  $(\phi,X)$ is replaced by $(\phi,X \setminus \{x\})$ and the size of the
  second component strictly decreases.  If splitting is successful and
  returns \(\{\phi_1,\dots ,\phi_n\}\) then $(\phi,X)$ is replaced by
  \(\{(\phi_i,SV(\phi_i, R))\}_{i=1\dots n}\). By
  Lemma~\ref{lem:MEASURE AFTER SPLITING DECREASES} the measures of
  goals from \(\{\phi_1,\dots ,\phi_n\}\) are strictly smaller than
  the measure of $\phi$ and consequently $M(W)$ strictly decreases.
\end{proof}

\begin{lem}\label{lem:CORRECTNESS OF THE ALGORITHM}
  If \(\Rew(\Gamma,R)\) preserves reducibility and the algorithm stops
  with $CE=\emptyset$ then the initial goal is covered.
\end{lem}
\begin{proof}
  Let us consider a successful run of the algorithm, performing a
  finite number of times the body of the \textrm{Repeat} loop and
  resulting in \(CE= \emptyset\). By induction on $n$, the number of
  \textrm{Repeat} steps until the end of the algorithm, we prove that
  the goals appearing in $W$ are covered.

  The base case, for \(n=0\), is trivial since $W_0$ is empty. 

  Now suppose that $n$ steps before the end of the algorithm all goals in
  $W_n$ are covered and let us check that this was true $n+1$ steps before the
  end, i.e.\ one step of the algorithm earlier. 
  
  In case $2$, $W_{n+1}$ contains all goals from $W_n$ and one goal $\phi$
  which is immediately covered by a rule in $R$. By preservation of
  reducibility (Lemma~\ref{IMMEDIATELY COVERED IS COVERED}) every
  normalized canonical instance of $\phi$ is also immediately covered
  and consequently all goals of $W_{n+1}$ are covered.

  Case $3(a)$ is impossible since it makes the set $CE$ non-empty.

  In case $3(b)\textit{i}$, $W_{n+1}$ contains some of the goals from $W_n$ and
  one goal $\phi$ whose subgoals resulting from successful splitting
  are already in $W_n$. By Lemma~\ref{NORMALIZED
    CANONICAL INSTANCES AFTER SPLITTING} the set of normalized
  canonical instances  of these subgoals contains the set of
  normalized canonical instances of~$\phi$. Hence $W_{n+1}$ is covered.

  In case $3(b)\textit{ii}$ the set of goals in $W_{n+1}$ and $W_n$
  are equal.

  Hence the initial goal in $W$ is also covered.
\end{proof}

\begin{exa}\label{ex:alg}
  The beginning of a possible run of the algorithm for the function
  \texttt{rotr} is presented already in
  Example~\ref{ex:splitting}. Both splitting operations are
  performed on safe variables, as required. We are left with the
  goal
  \texttt{rotr (S (nA + nC)) (Node nA A b nC C)}.
  Splitting along \texttt{A} results in:
  \begin{program}
rotr (S (O + nC)) (Node O Leaf b nC C)
rotr (S((S(nX+nZ))+nC)) (Node (S(nX+xZ)) (Node nX X y nZ Z) b nC C)
\end{program}
immediately covered by the second and the third rule respectively.

Since we started with the initial goal \texttt{rotr n t} and since
the definition of \texttt{rotr} preserves reducibility, it
is complete.

\end{exa}

\noindent
When the coverage checking algorithm stops with $CE \not = \emptyset$,
we cannot deduce that $R$ is complete. The set $CE$ contains potential
counterexamples.  They can be true counterexamples, false counterexamples, 
or goals for which splitting failed along all safe variables, due to
incompleteness of the unification algorithm. In some cases further
splitting of a false counterexample may result in reducible goals or in
the elimination of the goal as uninhabited, but it may also loop.
Some solutions preventing looping (finitary splitting) can be
found in~\cite{schurmann03coverage}.

Unfortunately splitting failure due to incompleteness of the
unification may happen while checking coverage of a definition by case
analysis over complex dependent inductive types (for example trees of size 2),
even if rules for all constructors are given.
Therefore, it is advisable to add a second phase to our algorithm,
which would treat undefined output of unification as success.  Using
this second phase of the algorithm, one can accept all definitions by
case analysis that can be written in Coq.

\begin{exa}\label{ex:glupi-split} Let \texttt{g} be
  a function defined by case analysis and let \texttt{g'} be its
  version defined by rewriting (without the impossible
  \texttt{Leaf} case):
\begin{program}
Definition g (t : tree (S (S O))) : bool := match t with
  Leaf \(\Rightarrow\) false
| Node _ Leaf _ _ _ \(\Rightarrow\) true
| Node _ (Node _ _ _ _ _) _ _ _ \(\Rightarrow\) false
end.

Symbol g' : tree (S (S O)) \ra bool
Rules 
  g' (Node ?1 Leaf b ?2 t') \lra true
  g' (Node ?1 (Node ?2 t b ?3 t') b' ?4 t'') \lra false
\end{program}
\end{exa}
Our algorithm starts with the goal \texttt{g' x}, splits it along
\texttt{x}, easily detects that \texttt{Leaf} case is not possible but
is stuck on \texttt{Node n t b m t'}, because this requires deciding
that \texttt{S (n+m)} is unifiable with \texttt{S (S O)}, which may be
too hard for a unification algorithm\footnote{Note that a better
  unification algorithm could find the two most general
  solutions   \texttt{n=O}, \texttt{m=(S O)} and \texttt{n=(S O)},
  \texttt{m=O}. Then splitting would result in two goals immediately
  covered by rules for \texttt{g'}.}. In that case the initial goal
\mbox{\texttt{g' x}} becomes a potential counterexample.

\paragraph{\bf Accepting all definitions by case analysis.}
The second
phase of our algorithm would start only for the goals with
safe splitting variables, i.e.\ where regular splitting failed because
the unification was too weak. In this phase, the splitting would
become lax by treating ?\ unification result as successful
and returning simple substitutions \(\sigma_c=\{x  \mapsto  c(\vec{z})\}\)
for such cases (see Definition~\ref{def:SPLITTING ALONG x}). As a result the
goals would not be well-typed sequents anymore, which has to be taken
into account by the unification algorithm. On the other hand
typability is not required for splitting matching and the rest of the
algorithm which would work just like described in Definition~\ref{def:COVER
  CHECKING ALGORITHM}.  Both arguments of termination and
correctness of the algorithm would hold.

Going back to our example. Redoing the lax splitting on \texttt{x} in
the goal \texttt{g' x}, one gets again that \texttt{Leaf} is
impossible, but \texttt{Node} is now accepted and leads to an
(untyped) goal \texttt{g' (Node n t b m t')}.  Splitting on \texttt{t}
is now successful for both constructors and both resulting goals get
reduced.

\expandafter\ifx\csname SourceFile\endcsname\relax\else\SourceFile{conclusions.ltx}\fi 
\section{More examples}



\subsection{Heterogeneous equality}
Consider the inductive predicate \texttt{JMeq} of heterogeneous
equality with its non-standard elimination rule:



\begin{program}
Inductive JMeq (A:Set)(a:A): forall B:Set, B \ra Set := JMrefl: JMeq A a A a.
  
Symbol JMelim : forall (A:Set)(a:A)(P: forall b:A, JMeq A a A b -> Set),
                P a (JMrefl A a) \ra forall (b: A) (e: JMeq A a A b), (P b e) 
Rules 
  JMelim A a P h a (JMrefl A a) \lra h
\end{program}
One splitting of \texttt{JMelim A a P h b c} over \texttt{c} results in
\texttt{JMelim A a P h a (JMrefl A a)} which is equal to the left-hand
side of the rule. Hence this rule completely defines \texttt{JMelim}.

\subsection{Uniqueness of Identity Proofs and Streicher's axiom K}
Consider the type \texttt{eq} and the definition of function \texttt{UIP}, 
proving that identity proofs are unique:
\begin{program}
Inductive  eq (A:Set)(a:A): A \ra Set  :=  refl: eq A a a.
  
Symbol UIP : forall (A:Set)(a b:A)(p q: eq A a b), (eq (eq A a b) p q)
Rules 
  UIP A a a (refl A a) (refl A a) \lra refl (eq A a a) (refl A a).
\end{program}
The function \texttt{UIP} is completely defined since two subsequent
splittings of \texttt{UIP A a b p q}, along \texttt{p} and along \texttt{q},
result in \texttt{UIP A a a (refl A a) (refl A a)} which is exactly the
left-hand side of the only rule for \texttt{UIP}. 

The rule for Streicher's axiom K can also easily be proved complete:
\begin{program}
Symbol K : forall (A:Set) (a:A) (P:eq A a a \ra Set),
   P (refl A a) \ra forall p: eq A a a, P p
Rules
   K A a P h (refl A a) \lra h
\end{program}
Note that both rules for \texttt{UIP} and \texttt{K} can also be written
in a left-linear form:
\begin{program}
   UIP A a ?1 (refl ?2 ?3) (refl ?4 ?5) \lra refl (eq A a a) (refl A a)
\medskip   K A a P h (refl ?1 ?2) \lra h
\end{program}

\subsection{Non pattern matching rules}

These are two examples of complete definitions which do not follow the
pattern matching schemes as defined in~\cite{DCoquand92typesa}
and~\cite{mcbride00dependently}. 

\begin{program}
Symbol or' : bool \ra bool \ra bool
  Rules
    or' x x \lra x
    or' true y \lra true    
    or' x true \lra true
\end{program}

\begin{program}
Symbol lt, diff : nat \ra nat \ra bool
  Rules
    lt O y \lra diff O y
    lt x O \lra false    
    lt (S x) (S y) \lra lt x y 

    diff x x \lra false
    diff O (S y) \lra true
    diff (S x) O \lra true
    diff (S x) (S y) \lra diff x y
\end{program}

\section{Conclusions and Related Work}
In this paper we study consistency of the calculus of constructions
with rewriting. More precisely, we propose a formal system extending
an arbitrary PTS with inductive definitions and definitions by
rewriting. Assuming that suitable positivity and acceptance
conditions guarantee termination and confluence, we formalize the
notion of a complete definition by rewriting.  We show that in every
environment consisting only of inductive definitions and complete
definitions by rewriting there is no proof of $\prd{x}{*}x$.
Moreover, we present a sound and terminating algorithm for checking
completeness of definitions. It is necessarily incomplete, since  
in presence of dependent types emptiness of types trivially reduces to
completeness and the former is undecidable.

Our coverage checking algorithm resembles the one proposed by Coquand
in~\cite{DCoquand92typesa} for Martin-L\"of type theory and used by
McBride for his OLEG calculus~\cite{mcbride00dependently}. In these
works the procedure consisting in successive case-splittings is used
to interactively built pattern matching equations, or to check that a
given set of equations can be built this way. Unlike in our paper,
Coquand and McBride do not have to worry whether all instances 
of a reducible subgoal are reducible.
Indeed, in~\cite{DCoquand92typesa}
pattern matching equations are meant to be applied to terms modulo
conversion, and in~\cite{mcbride00dependently} equations (or rather
the order of splittings in the successful run of the coverage checking
procedure) serve as a guideline to construct an OLEG term verifying
the equations.  Equations themselves are never
used for reduction and the constructed term reduces according to
existing rules. 

In our paper rewrite rules are matched against terms modulo $\alpha$-conversion.
Rewriting has to be confluent, strongly normalizing and has to
preserve reducibility. Under these assumptions we can prove
completeness for all examples from~\cite{DCoquand92typesa} and for the
class of pattern matching equations considered
in~\cite{mcbride00dependently}.  In particular we can deal with
elimination rules for inductive types and with Streicher's axiom K.  
Moreover, we can accept definitions which depart from standard pattern
matching, like \texttt{rotr} and $+$.

The formal presentation of our algorithm is directly inspired by the
work of Pfenning and Sch\"urmann~\cite{schurmann03coverage}.
A motivation for that paper was to verify that a logic program in the
Twelf prover covers all possible cases. In LF, the base calculus of
Twelf, there is no polymorphism, no rewriting and conversion is modulo
$\beta\eta$-conversion.  The authors use higher-order matching modulo~$\beta\eta$-conversion, 
which is decidable for patterns a la Miller and strict
patterns. Moreover, since all types and function symbols are known in
advance, the coverage is checked with respect to all available
function symbols. In our paper, conversion contains rewriting and it
cannot be used for matching; instead we use matching modulo
$\alpha$. This simplifies the algorithm searching for safe
splitting variables, but on the other hand it does not fit well with
instantiation and normalization. To overcome this problem we introduce
the notions of normalized canonical instance and preservation of
reducibility which were not present in previously mentioned papers.
Finally, since the sets of function symbols and rewrite rules grow as
the environment extends, coverage is checked with respect to constructors
only.

Even though the worst-case complexity of the coverage checking is
clearly exponential, for practical examples the algorithm should be
quite efficient. It is very similar in spirit to the algorithms
checking exhaustiveness of definitions by pattern matching in
functional programming languages and these are known to work
effectively in practice.

An important issue which is not addressed in this paper is to know how
much we extend conversion. Of course it depends on the choice of
conditions \Acc\ and \Pos\ and on the unification algorithm used for
coverage checking. In particular, some of the definitions by pattern
matching can be encoded by recursors~\cite{cornes:phd}, so if \Acc\ is
strict, we may have no extension at all. In general there seems to be
at least two kinds of extensions. The first are non-standard
elimination rules for inductive types, but the work of McBride shows
that the axiom K is sufficient to encode all other definitions by
pattern matching considered by Coquand. The second are additional
rules which extend a definition by pattern matching (like
associativity for $+$). It is known that for first-order rewriting,
these rules are inductive consequences of the pattern matching ones,
i.e.\ all their canonical instances are satisfied as equations
(see e.g.\ Theorem~7.6.5 in~\cite{terese}). Unfortunately, this is no longer true
for higher-order rules over inductive types with functional arguments.
Nevertheless it seems that such rules are inductive consequences of
the pattern matching rules if the corresponding equality is
extensional.

Finally, our completeness condition \Comp\ verifies closure properties
defined in \cite{chrzaszcz03types,chrzaszcz04phd}. Hence, it is
adequate for a smooth integration of rewriting with the module system
present in Coq since its version 7.4.


\section*{Acknowledgement}
The authors wish to thank Pawe\l\ Urzyczyn and anonymous referees for
their helpful comments and suggestions.

\bibliographystyle{plain}
\bibliography{abbrevs,demons,demons2,demons3,crossrefs,crossrefs2,daria,bib}

\end{document}